# Methods for fast and accurate material properties estimate with terahertz time-domain spectroscopy in transmission and reflection with optically thick materials

VINCENT GOUMARRE[1,†,*], YASITH AMARASINGHE[1,†], MARTIN LAHN HENRIKSEN[2], MOGENS HINGE[2], PERNILLE KLARSKOV[1]

[1] *Terahertz Photonics, Department of Electrical and Computer Engineering, Aarhus University, Finlandsgade 22, DK-8200, Aarhus N, Denmark*
[2] *Plastic and Polymer Engineering, Department of Biological and Chemical Engineering, Aarhus University, Aabogade 40, DK-8200 Aarhus N., Denmark*
[†] *These authors contributed equally to this work*
*\*vincent.goumarre@ece.au.dk*

**Abstract:** With the development of terahertz time-domain spectroscopy, methods have been proposed to precisely estimate the thickness, refractive index, and attenuation coefficient of a sample. In this article, we propose a new method to compute these parameters. In this method, the attenuation is expressed in function of the refractive index. The theoretical unwrapped angle, which therefore only depends on the refractive index, is then matched to the experimental value. By applying already existing methods for estimating the thickness of an optically thick sample, the dielectric properties of the sample can be deduced. The method is applied both to the transmission and reflection spectroscopy. A demonstration of the method and a comparison with the previous methods are finally shown.

## 1. Introduction

Emerging at the end of the 20[th] century [1, 2], terahertz time-domain spectroscopy has been applied in many domains for material characterization. Its non-destructive nature makes it a good technique to measure dielectric properties and has been applied, among other things, to semiconductors [3], wood [4], ceramics [5], paintings [6] and plastics [7].

The typical design of the experiment consists of a terahertz emitter and receiver placed on both sides of a sample with parallel sides (so called transmission spectroscopy). To characterize a material's properties correctly, time domain spectroscopy must rely on accurate methods to estimate the refractive index and absorption coefficient. To do so, several methods have been proposed and different reviews try to list these attempts [8, 9]. Some methods rely on a time-domain analysis [10] or the Kramers-Kronig analysis [11]. The main method for the computation of the dielectric properties remains the method developed by Dorney et al. [12]. It is based on a first rough estimate of the material properties, which are then iteratively corrected using the Fabry-Perot effect. For thick samples, it has been shown that the material thickness can be precisely estimated using the Fabry-Perot effect [12-14]. Knowledge of the thickness is essential to an accurate retrieval of the material parameters. A bad estimate of the thickness implies a shift of the refractive index and absorption coefficient and the appearance of wavelets that can be used to identify a thickness misestimate. For thin samples, because the effect of the Fabry-Perot is spanned over several terahertz, the wavelets cannot be used to estimate the thickness. In this case, other methods are applied for the thickness estimate such as the definition of a quasi-space [15] or the fitting of the refractive index with a function without inflection point [16].

Although reflection spectroscopy might be more adapted for industrial applications such as the measurement on an assembly line with a conveyer belt, its theoretical grounds have been less developed. Most of the parameter reflection spectrometry is done with a liquid behind a window [17-20] or for multilayer painting on a substrate [6, 21, 22]. These applications use either a Debye model on the permittivity, time-windowing of the sample signal, or even the prior knowledge of the material properties to deduce its thickness. Recently, a measure of substrate thickness and dielectric properties were proposed, where the substrate lies on a mirror [23]. Once again, a time-window was applied to the incoming signal to separate the signal reflected at the interface air-substrate and substrate-mirror.

This paper proposes to improve both the transmission and reflection spectroscopy computations by introducing a new method. This method does not rely on an iterative algorithm, but on the resolution of an equation containing a single parameter (the refractive index). This method considers the Fabry-Perot effect to improve the thickness estimates, both in transmission and reflection.

## 2. Notations and conventions

The complex refractive index is noted:

$$\tilde{n} = n - j\kappa, \tag{1}$$

with $\kappa$ being the attenuation coefficient. The absorption coefficient is defined by:

$$\alpha = \kappa \frac{2\omega}{c}, \tag{2}$$

where $\omega = 2\pi f$ is the angular frequency and $c$ the speed of light.

The reflection and transmission coefficients throughout this study are the Fresnel coefficient for the electric field [18]. Because the sample has a low attenuation coefficient ($\kappa \ll n$), the Fresnel coefficients are considered real:

$$T_{12} = \begin{cases} \dfrac{2\, n_1 \cos\theta_1}{n_2 \cos\theta_1 + n_1 \cos\theta_2} & \text{for p polarization,} \\ \dfrac{2\, n_1 \cos\theta_1}{n_1 \cos\theta_1 + n_2 \cos\theta_2} & \text{for s polarization,} \end{cases} \tag{3}$$

$$R_{12} = \begin{cases} \dfrac{n_1 \cos\theta_2 - n_2 \cos\theta_1}{n_2 \cos\theta_1 + n_1 \cos\theta_2} & \text{for p polarization,} \\ \dfrac{n_1 \cos\theta_1 - n_2 \cos\theta_2}{n_1 \cos\theta_1 + n_2 \cos\theta_2} & \text{for s polarization.} \end{cases} \tag{4}$$

Where 1 and 2 the incident and transmitted material and the angles are determined by Snell's law:

$$n_1 \sin\theta_1 = n_2 \sin\theta_2. \tag{5}$$

There are three materials used in this study:

- The sample (subscript $s$) to characterize.
- Dried air (subscript $a$) with a refractive index approximated to $n_a = 1$.
- A fully reflective mirror (subscript $m$) with reflection coefficient $R_{am} = -1$.

The sample used is supposed to be thick enough to separate temporally the signal from the different Fabry-Perot reflections. Such a sample is defined as optically thick [24]. Moreover, it is supposed to have a refractive index low enough, leading to the following relation:

$$T_{as} T_{sa} \gg |R_{sa}|. \tag{6}$$

In practice, the methods developed later work at least until $n \approx 3.5$, as demonstrated by the application on a silicon wafer (see Section 5).

Because the study is done in transmission and in transmission, the superscript denotes the case of the transmission ($t$) or the reflection ($r$).

The complex number are noted using the following convention:

$$Z = |Z|e^{-j\Phi_Z} = |Z|(\cos \Phi_Z - j \sin \Phi_Z). \tag{7}$$

## 3. Transmission

### 3.1 Model

The model for the transmission geometry is shown in Fig. 1.

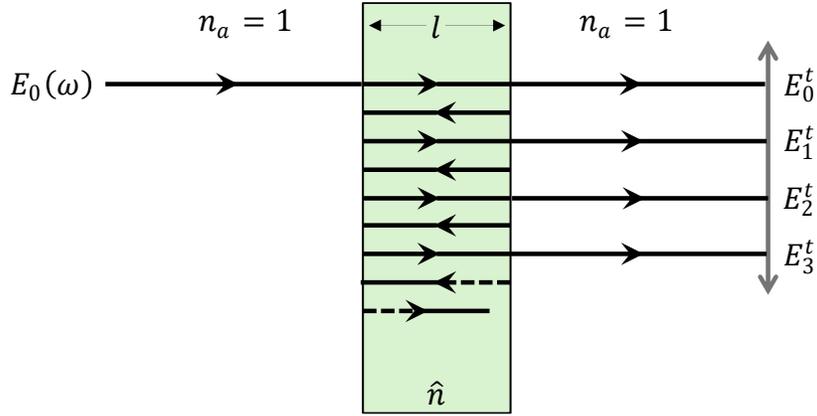

Fig. 1. Wave propagation for the transmission configuration.

It has been shown that the total transmission transfer function, defined by the ratio of the sample by reference spectra, is [25]:

$$H^t(\omega) = \frac{E^t_{sam}(\omega)}{E^t_{ref}(\omega)} = \frac{T^t_{as} T^t_{sa} e^{-\kappa \frac{\omega l}{c}} e^{-j(n-1)\frac{\omega l}{c}}}{1 - R^{t}_{sa}{}^2 e^{-2\kappa \frac{\omega l}{c}} e^{-2jn\frac{\omega l}{c}}}. \tag{8}$$

With

$$T^t_{as} = \frac{2}{n+1}, \quad T^t_{sa} = \frac{2n}{n+1}, \text{ and } R^t_{sa} = -\frac{n-1}{n+1} \tag{9}$$

are the Fresnel coefficients introduced in Eqs. (3) and (4) with $\theta_1 = \theta_2 = 0$.

### 3.2 Signal processing

Eq. (8) depends on the experimental value of the transfer function on the left-hand side and two material dielectric parameters on the right-hand side: the refractive index $n$ and the attenuation coefficient $\kappa$. The goal is to find an equation that depends on the refractive index alone. Using the unwrapped phase [26] and modulus of Eq. (8), the following set of equations can be deduced:

$$\angle H^t(\omega) = (n-1)\frac{\omega l}{c} \\ + atan2\left[R^{t}_{sa}{}^2 e^{-2\kappa\frac{\omega l}{c}} \sin\left(2n\frac{\omega l}{c}\right), 1 - R^{t}_{sa}{}^2 e^{-2\kappa\frac{\omega l}{c}} \cos\left(2n\frac{\omega l}{c}\right)\right], \tag{10}$$

$$R_{sa}^t{}^4 e^{-4\kappa\frac{\omega l}{c}} - \left[\frac{T_{as}^t{}^2 T_{sa}^t{}^2}{|H|^2} + 2R_{sa}^t{}^2 \cos\left(2n\frac{\omega l}{c}\right)\right] e^{-2\kappa\frac{\omega l}{c}} + 1 = 0. \tag{11}$$

Eq. (10) is true as long as $atan2$ is continuous and must therefore not be unwrapped. Mathematically, this happens if $R_{sa}^t{}^2 e^{-2\kappa\frac{\omega l}{c}} < 1$, which is theoretically satisfied, since $|R_{sa}^t| < 1$ and $e^{-2\kappa\frac{\omega l}{c}} < 1$.

Eq. (11) is the root of a second-degree polynomial on $e^{-2\kappa\frac{\omega l}{c}}$ leading to the following determinant:

$$\Delta(n) = \left[\frac{T_{as}^t{}^2 T_{sa}^t{}^2}{|H|^2} + 2R_{sa}^t{}^2 \cos\left(2n\frac{\omega l}{c}\right)\right]^2 - 4R_{sa}^t{}^4. \tag{12}$$

This discriminant changes sign only when $\frac{T_{as}^t{}^2 T_{sa}^t{}^2}{|H|^2} + 2R_{sa}^t{}^2 \cos\left(2n\frac{\omega l}{c}\right) = 2R_{sa}^t{}^2$.

The solution for the second-degree polynomial in the region $\Delta(n) > 0$ are:

$$\left(e^{-2\kappa\frac{\omega l}{c}}\right)_n^{\pm} = \frac{\frac{T_{as}^t{}^2 T_{sa}^t{}^2}{|H|^2} + 2R_{sa}^t{}^2 \cos\left(2n\frac{\omega l}{c}\right) \pm \sqrt{\Delta(n)}}{2R_{sa}^t{}^4}. \tag{13}$$

Both solutions are positive but the requirement $e^{-2\kappa\frac{\omega l}{c}} \lesssim 1$ must also be satisfied. By applying the approximation described in Eq. (6), the two solutions are approximated by:

$$\left(e^{-2\kappa\frac{\omega l}{c}}\right)_n^{+} \approx \frac{T_{as}^t{}^2 T_{sa}^t{}^2/|H|^2 + 2R_{sa}^t{}^2 \cos(2n\,\omega l/c)}{R_{sa}^t{}^4} \gg 1,$$

$$\left(e^{-2\kappa\frac{\omega l}{c}}\right)_n^{-} \approx \frac{1}{T_{as}^t{}^2 T_{sa}^t{}^2/|H|^2 + 2R_{sa}^t{}^2 \cos(2n\,\omega l/c)} \lesssim 1.$$

Therefore, we define the following continuous function:

$$\left(e^{-2\kappa\frac{\omega l}{c}}\right)_n = \begin{cases} \left(e^{-2\kappa\frac{\omega l}{c}}\right)_n^{-} & \text{if } \Delta(n) \geq 0, \\ 1/R_{sa}^t{}^2 & \text{if } \Delta(n) < 0. \end{cases} \tag{14}$$

In addition to the continuity of the function, the choice of the function in the case of a negative determinant also exhibits an interesting behavior when the angle difference $\Delta\Phi^t$ is define:

$$\Delta\Phi^t(n) = \angle H^t(\omega) - (n-1)\frac{\omega l}{c}$$
$$- atan2\left[R_{sa}^t{}^2 \left(e^{-2\kappa\frac{\omega l}{c}}\right)_n \sin\left(2n\frac{\omega l}{c}\right), 1 - R_{sa}^t{}^2 \left(e^{-2\kappa\frac{\omega l}{c}}\right)_n \cos\left(2n\frac{\omega l}{c}\right)\right]. \tag{15}$$

It can be shown that $\Delta\Phi^t$ remains constant when the discriminant is negative. When the discriminant is positive, the trend of $\Delta\Phi^t$ is most likely imposed by the term $(n-1)\,\omega l/c$, meaning that the function is always decreasing.

Eq. (10) imposes that the refractive index is defined such that $\Delta\Phi^t(n_0) = 0$. Because the $atan2$ function is in the range $[-\pi, \pi]$, we can define:

$$n_{min} = 1 + \frac{\angle H(\omega) - \pi}{\omega l/c}, \qquad (16)$$

$$n_{max} = 1 + \frac{\angle H(\omega) + \pi}{\omega l/c}, \qquad (17)$$

which correspond to the bounds to find $n_0$ since $\Delta\Phi^t(n_{min}) \geq 0$ and $\Delta\Phi^t(n_{max}) \leq 0$. From $x_{min}$ and $x_{max}$, the value of $n_0$ can be derived by using Dekker's algorithm [27].

From this value of the refractive index, the attenuation coefficient is simply be computed by:

$$\kappa_0 = - \frac{\log\left[\left(e^{-2\kappa\frac{\omega l}{c}}\right)^{-}_{n_0}\right]}{2\,\omega l/c}. \qquad (18)$$

## 4. Reflection

### 4.1 Model

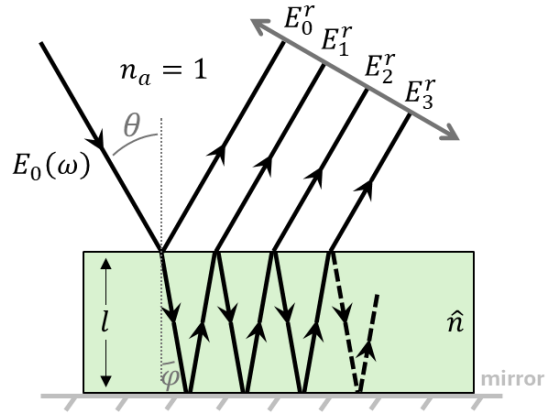

Fig. 2. Wave propagation for the reflection configuration.

The experimental setup for the reflection measurement is done so that the sample is placed on a mirror with reflection coefficient -1. The first measurement is done without the sample to get a reference. Then the sample is placed on the mirror. The goal will be to retrieve the material dielectric properties following a very similar procedure as the transmission spectroscopy by expressing the attenuation coefficient in terms of the refractive index and deducing an equation on the refractive index.

For the experiment shown in Fig. 2, the sample electric field can be expressed as [25]:

$$E^r_{sam}(\omega) = E_0(\omega) \cdot \left( R^r_{as} + \frac{T^r_{as} T^r_{sa} R_{sm}\, e^{-2\frac{\kappa}{\cos\varphi}\frac{\omega l}{c}} e^{-2jn\frac{\omega l \cos\varphi}{c}}}{1 - R_{sm} R^r_{sa} e^{-2\frac{\kappa}{\cos\varphi}\frac{\omega l}{c}} e^{-2jn\frac{\omega l \cos\varphi}{c}}} \right), \qquad (19)$$

with $\cos\varphi = \sqrt{1 - \sin^2\theta/n^2}$ and $\theta$ the experimental incidence angle.

Because the sample is placed on the mirror and might not be completely flat, a thin layer of air may remain under the sample. Even though the layer of air is generally small compared to the size of the sample (see Section 5.2), it can impact dramatically the shape of the refractive index, as shown in Fig. 3.

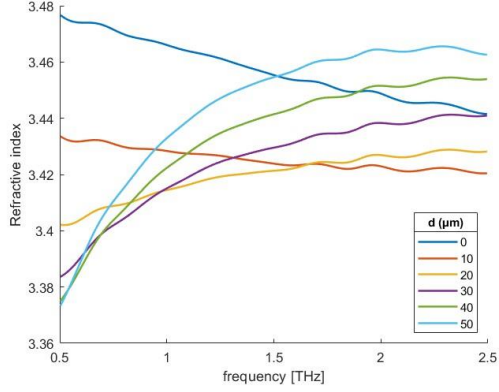

Fig. 3. Influence of the layer of air d on the refractive index of a 521 μm thick silicon wafer.

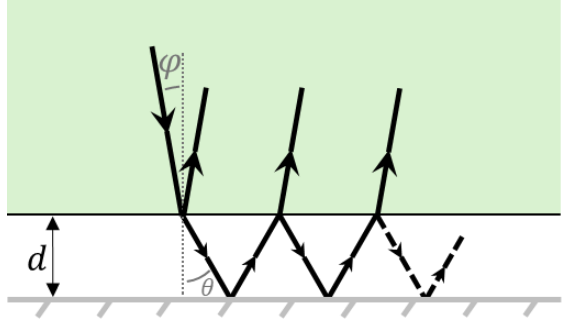

Fig. 4. Wave propagation at the interface between sample and mirror for the reflection configuration.

The interaction between the sample and the mirror including a layer of air is depicted in Fig. 4. This does not change the form of Eq. (9), but influences the value of $R_{sm}$. If the sample was perfectly lying on the mirror, the reflection coefficient would be $R_{sm} = R_{am} = -1$. However, the layer of air induces a correction to this value:

$$R_{sm} = H_{sm} R_{am} e^{-2j\frac{\omega d \cos\theta}{c}}, \tag{20}$$

with

$$H_{sm} = \frac{T_{sa}^r T_{as}^r}{1 - R_{as}^r R_{am} e^{-2j\cos\theta \frac{\omega d}{c}}} + \frac{R_{sa}^r}{R_{am}} e^{2j\frac{\omega d \cos\theta}{c}} = |H_{sm}| e^{-j\Phi_{sm}}. \tag{21}$$

It can be shown that in the presence of a perfectly reflecting mirror ($R_{am} = -1$), the modulus $|H_{sm}|$ is 1. This simply means that the light ray is completely sent back in the sample but introducing a phase shift. It can be shown that the phase of $H_{sm}$ is expressed as:

$$\Phi_{sm} = -atan2\left[\left(\frac{R_{sa}^r}{R_{am}} - R_{as}^r R_{am}\right)\sin\left(2\cos\theta\,\frac{\omega d}{c}\right) - R_{as}^r R_{sa}^r \sin\left(4\cos\theta\,\frac{\omega d}{c}\right),\right.$$
$$\left. 1 + \left(\frac{R_{sa}^r}{R_{am}} - R_{as}^r R_{am}\right)\cos\left(2\cos\theta\,\frac{\omega d}{c}\right) - R_{as}^r R_{sa}^r \cos\left(4\cos\theta\,\frac{\omega d}{c}\right)\right]. \tag{22}$$

The reference signal, measured by removing the sample from the mirror is given by:

$$E_{ref}^r(\omega) = E_0(\omega) \cdot R_{am} e^{-2j\frac{\omega(l+d)\cos\theta}{c}}. \tag{23}$$

From Eqs. (19), (20), (21) and (23) the transfer function becomes:

$$H^r(\omega) = \frac{E^r_{sam}(\omega)}{E^r_{ref}(\omega)} = H^r_{main}(\omega) + H^r_{echo}(\omega), \tag{24}$$

with

$$H^r_{main}(\omega) = R^r_{as} e^{2j\frac{\omega(l+d)\cos\theta}{c}}, \tag{25}$$

$$H^r_{echo}(\omega) = \frac{T^r_{as} T^r_{sa} H_{sm} e^{-2\frac{\kappa}{\cos\varphi}\frac{\omega l}{c}} e^{-2j\left(\sqrt{n^2-\sin^2\theta}-\cos\theta\right)\frac{\omega l}{c}}}{1 - H_{sm} R_{am} R^r_{sa} e^{-2\frac{\kappa}{\cos\varphi}\frac{\omega l}{c}} e^{-2j\left(\sqrt{n^2-\sin^2\theta}\frac{\omega l}{c}+\cos\theta\frac{\omega d}{c}\right)}} \tag{26}$$

$$= |H^r_{echo}| e^{-i\Phi_{H^r_{echo}}}.$$

In the current work, the main signal is removed from the sample signal. This is done by suppressing the peak appearing before the reference peak in the time domain. Two reasons motivate the removal of the main signal. First, the computation of the refractive index will be easier without the additive term $H^r_{main}$. More importantly, the mirror size for the terahertz emitter and receiver in our experiment is the same. This implies that, when the optical system is set up without the sample, the introduction of the sample modifies the path of the terahertz light ray the main signal is not caught fully by the receiver. This artificially modifies the apparent refractive index computed by the reflection Fresnel coefficient.

Similarly to the transmission case, we can deduce the following two equations from Eq. (26):

$$\angle H^r_{echo}(\omega) = 2\left(\sqrt{n^2-\sin^2\theta}-\cos\theta\right)\frac{\omega l}{c} + \Phi_{sm}$$
$$+\mathrm{atan2}\left[R^r_{sa} R_{am} |H_{sm}| e^{-2\frac{\kappa}{\cos\varphi}\frac{\omega l}{c}}\sin\left(2\sqrt{n^2-\sin^2\theta}\frac{\omega l}{c} + 2\cos\theta\frac{\omega d}{c} + \Phi_{sm}\right), \tag{27}\right.$$
$$\left. 1 - R^r_{sa} R_{am} |H_{sm}| e^{-2\frac{\kappa}{\cos\varphi}\frac{\omega l}{c}}\cos\left(2\sqrt{n^2-\sin^2\theta}\frac{\omega l}{c} + 2\cos\theta\frac{\omega d}{c} + \Phi_{sm}\right)\right].$$

$$\left(e^{-2\frac{\kappa}{\cos\varphi}\frac{\omega l}{c}}\right)_n = |H_{sm}|^{-1}\left[\frac{(T^r_{as} T^r_{sa})^2}{|H^r_{echo}|^2} + (R^r_{sa} R^r_{am})^2\right.$$
$$\left. +2\frac{T^r_{as} T^r_{sa} R^r_{sa} R^r_{am}}{|H^r_{echo}|}\cos\left(2\cos\theta\frac{\omega(l+d)}{c} + \Phi_{H^r_{echo}}\right)\right]^{-\frac{1}{2}}. \tag{28}$$

By replacing $e^{-2\frac{\kappa}{\cos\varphi}\frac{\omega l}{c}}$ in Eq. (27) with its value in Eq. (28), one can find an equation that only depends on the refractive index and experimental values:

$$\Delta\Phi^r(n) = \angle H^r_{echo}(\omega) - 2\left(\sqrt{n^2-\sin^2\theta}-\cos\theta\right)\frac{\omega l}{c} - \Phi_{sm}$$
$$-\mathrm{atan2}\left[R^r_{sa} R_{am} |H_{sm}|\left(e^{-2\frac{\kappa}{\cos\varphi}\frac{\omega l}{c}}\right)_n \sin\left(2\sqrt{n^2-\sin^2\theta}\frac{\omega l}{c} + 2\cos\theta\frac{\omega d}{c} + \Phi_{sm}\right), \tag{29}\right.$$
$$\left. 1 - R^r_{sa} R_{am} |H_{sm}|\left(e^{-2\frac{\kappa}{\cos\varphi}\frac{\omega l}{c}}\right)_n \cos\left(2\sqrt{n^2-\sin^2\theta}\frac{\omega l}{c} + 2\cos\theta\frac{\omega d}{c} + \Phi_{sm}\right)\right].$$

This function is more likely to be decreasing due to the dominance of the $2\sqrt{n^2-\sin^2\theta}\frac{\omega l}{c}$ term. The solution $\Delta\Phi^r(n_0) = 0$ is comprised between the two following values:

$$n_{min} = \sqrt{\left(\frac{\angle H^r(\omega) - 2\pi}{2\omega l/c} + \cos\theta\right)^2 + \sin^2\theta}, \tag{30}$$

$$n_{max} = \sqrt{\left(\frac{\angle H^r(\omega) + 2\pi}{2\omega l/c} + \cos\theta\right)^2 + \sin^2\theta}. \tag{31}$$

Similarly to the transmission method, the equation is solved by the Dekker's method to find the root [27], and the attenuation coefficient is defined as:

$$\kappa_0 = -\log\left[\left(e^{-2\frac{\kappa}{\cos\varphi}\frac{\omega l}{c}}\right)_{n_0}\right]\frac{\sqrt{1 - \sin^2\theta/n_0^2}}{2\,\omega l/c} \tag{32}$$

### 4.2 Thickness determination

Eq. (26) implies that two thicknesses must be deduced in the reflection geometry: $l$ and $d$. Because the sample is optically thick, its thickness can be treated as done in Ref. [12], where a total variation is defined by:

$$TV = \sum_k \left(|n(\omega_k) - n(\omega_{k-1})| + |\kappa(\omega_k) - \kappa(\omega_{k-1})|\right). \tag{33}$$

This total variation must then be minimized by adjusting the thickness by use of a Nelder-Mead algorithm.

Experimentally, as shown in Fig. 5, this process is only sensitive to the $l + d$ value, meaning that it cannot be used to estimate the thickness of the layer of air under the sample.

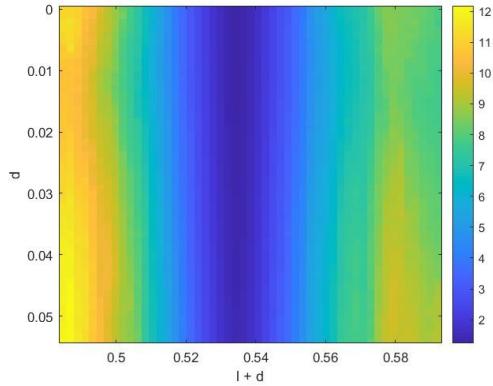

Fig. 5. Total variation depending on the layer of air and sample thicknesses of a 521 μm thick silicon wafer.

The layer of air under the sample is more likely to be a thin sample. In our experiments, the samples have a thickness typically between 0.5 and 5 mm, and the layer of air is ~0.01 mm. Because the layer of air modifies the refractive index shape by introducing inflection (see Fig. 3), its thickness is determined by the method developed in Ref. [16]. Once $l + d$ is fixed, the refractive index is fitted by a function:

$$f_{fit}(\omega) = a_0 e^{a_1 \omega} + a_2 \omega + a_3, \tag{34}$$

where $a_k$ are the variables to be adjusted by the fitting process. The sum of squares due to error is then deduced. Minimizing this error gives the layer of air thickness $d$.

## 5. Application to a Silicon wafer

The methods presented in this paper are applied to the measurement of the dielectric properties of a silicon wafer with a thickness given by the manufacturer at $525 \pm 20$ µm and measured with a micrometer to 524 µm. The Terahertz system used is a TOPTICA TeraFlash Pro generating terahertz impulses with a spectral range between 0.1 and 6 THz.

### 5.1 Transmission

In the transmission geometry, the terahertz ray is focused on the sample thanks to off-axis parabolic mirrors. The transmission data is treated using with three different methods:

- Without any optimization ("unoptimized"): The sample thickness is computed by detecting the sample and reference peaks applying the following formula [28]:

$$l = \frac{c(t_{echo1} - t_{echo0}) - 2c(t_{echo0} - t_{ref})}{2}, \quad (35)$$

  with $t_{echo1}$, $t_{echo0}$ and $t_{ref}$ are the time of arrival of $E_1^t$, $E_0^t$ and $E_{ref}$ respectively. The different times are taken at the minimum value of the pulse.
  The refractive index and the attenuation coefficient are computed by isolating the reference and first sample peaks thanks to a Tukey window [29] and applying the following formulas [25]:

$$n = 1 + \frac{\Phi_{echo0} - \Phi_{ref}}{\omega l/c}, \quad (36)$$

$$\kappa = \frac{\ln(T_{as}^t T_{sa}^t) - \ln(|E_{echo0}|/|E_{ref}|)}{\omega l/c}. \quad (37)$$

- Dorney's method: Developed by Dorney et al. [12], allows to compute more precisely the material parameters, including the thickness by using the Melder-Nead method. The initial thickness, refractive index and attenuation coefficient are the ones from the unoptimized case.
- The method developed in Section 3.

The results of these methods are shown in Fig. 6 and Table 1.

Table 1. Width of the material and computing time given by the different methods.

| Method | Width (µm) | Computation time (s) |
| --- | --- | --- |
| micrometer | 524 | - |
| Unoptimized | 524.64 | ~0.3 |
| Dorney | 521.44 | ~170 |
| This work | 521.41 | ~1 |

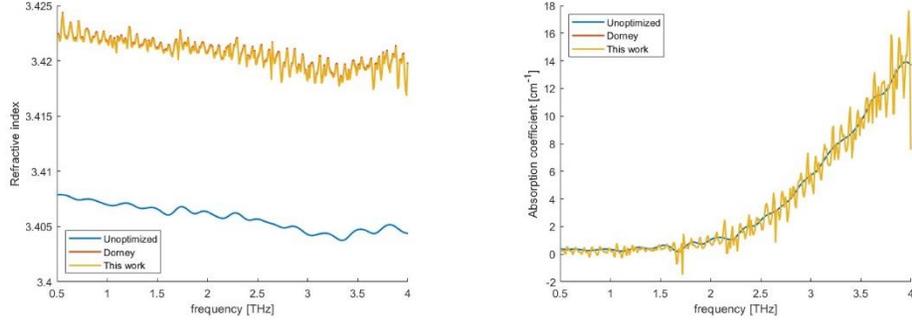

Fig. 6: Refractive index (left) and attenuation coefficient (right) for the unoptimized, Dorney's and this work's methods.

From Fig. 6, the refractive index computed by Dorney and by the method presented in Section 3 are almost identical. The only difference comes from a negligible difference in the width estimate. However, the computing time has been divided by a factor ~100. This is due to several factors. First, Dorney's method is based on an iterative algorithm while the algorithm developed earlier only requires finding the root of a function given two points with different signs. Moreover, Dorney's method requires complex computation that are computationally heavy while the $\Delta\Phi^t$ function that must be set to zero only handles real numbers.

### 5.2 Reflection

For the reflection geometry, a TOPTICA compact optical head is used using off-axis parabolic mirrors to focus the beam on a polished aluminum mirror. The angle of incidence of the setup is $\theta = 8.8°$. Two computational methods are used to deduce the material properties:

- Without any optimization: The sample thickness is estimated using the reference, $E_1^r$ and $E_2^r$ signals arrival time:

$$l = \frac{c(t_{echo2} - t_{echo1}) - c(t_{echo1} - t_{ref})}{2 \cos \theta}. \tag{38}$$

This does not consider the possible layer of air between the sample and the mirror ($d = 0$). The refractive index and the attenuation coefficient are computed by isolating $E_1^r$ and the reference signal and proceeding to a Fourier transform:

$$n = \sqrt{\left(\frac{\Phi_{echo} - \Phi_{ref}}{2\,\omega l/c} + \cos\theta\right)^2 + \sin^2\theta}, \tag{39}$$

$$\kappa = \frac{\sqrt{1 - \sin^2\theta} \cdot \left(\ln(T_{as}^r T_{sa}^r) - \ln(|E_{echo}|/|E_{ref}|)\right)}{2\,\omega l/c}. \tag{40}$$

- The method developed in Section 4.

The results are compared to the measurements presented earlier in transmission and are shown in Fig. 7 and Table 2.

Table 2: Width of the material and computing time from the different methods.

| Method | Width (μm) | Air layer (μm) | Computation time (s) |
| --- | --- | --- | --- |
| micrometer | 524 | - | - |
| Unoptimized | 538.47 | 0 | ~0.5 |
| Transmission | 521.41 | - | ~1 |
| This work | 521.23 | 13.45 | ~20 |

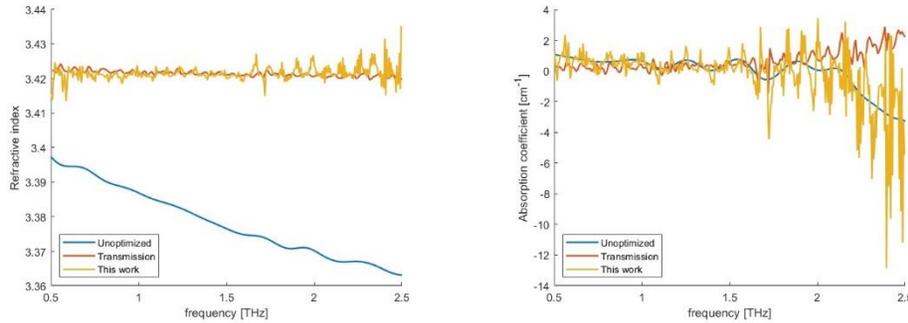

Fig. 7: Refractive index (left) and absorption coefficient (right) for the unoptimized and this work's methods in reflection, as well as the transmission measurement presented in Section 5.1.

The main contribution of the method on the sample properties measurements occurs for the refractive index. Even though the layer of air is estimated to approximately 2% of the wafer thickness, the consequence on the refractive index is important, as shown in Fig. 7. By estimating the correct layer of air thickness, the transmission and reflection measurement are matching, both for refractive index and for sample thickness (see Table 2). Concerning the attenuation coefficient, the values in transmission and in reflection match except for high frequencies, where the transmission method estimates an increasing absorption coefficient while it is decreasing for reflection. However, the amount of noise suggests that the absorption coefficient at high frequency is not reliable due to a signal over noise ratio being too small.

## 6. Conclusions

In this paper, we developed a new method to precisely compute the refractive index and the attenuation coefficient. On the contrary to the previously existing methods, this is built upon the resolution of the full transfer function, including the Fabry-Perot effect. This relies on the fact that the unwrapped phase of the transfer function implies a single refractive index as a solution [30].

The originality of this method is to express the attenuation coefficient in terms of the refractive index, thus implying the resolution of an equation depending on a single variable. This method allows both a very precise and fast estimate of the material thickness, refractive index, and absorption coefficient. As a proof of concept, the method is successfully applied on a Silicon wafer and can be applied to other samples with parallel interfaces.

Moreover, a similar technique is applied to reflection geometry. The unwrapped angle is matched to the experimental value by adjusting the refractive index. The absorption coefficient is deduced from the refractive index. The experiment consists of a sample put onto a mirror, with the possibility that a layer of air between the sample and the mirror remains. Therefore, the thickness estimate must be done twice, using techniques adapted for optically thick (sample) and thin (air) samples. The method is adapted for industrial settings where precision and speed are required.

### Acknowledgment

The authors thank the Innovation Fond Denmark (grant no: 0177-00035B) Aarhus University, Dansk Affaldsminimering Vestforbrænding, and Plastix for financial support within the Re-Plast project.